\documentclass[epj]{svjour}
\usepackage{latexsym}
\usepackage{graphics}

\begin{document}

\title{Search for electron antineutrino interactions with the Borexino 
Counting Test Facility at Gran Sasso}

\author{M.~Balata\inst{1}, G.~Bellini\inst{2}, J.~Benziger\inst{3},
 S.~Bonetti\inst{2}, B.~Caccianiga\inst{2}, F.~Calaprice\inst{4}, D.~D'Angelo\inst{5}, 
 A~de~Bellefon\inst{6}, H.~de~Kerret\inst{6}, A.~Derbin\inst{7},
A.~Etenko\inst{8}, R.~Ford\inst{4}, D.~Franco\inst{9}, C.~Galbiati\inst{4}, S.~Gazzana\inst{1},
M.~Giammarchi\inst{2}, A.~Goretti\inst{1}, E.~Harding\inst{4},
G.~Heusser\inst{9}, A.~Ianni\inst{1}\mail{aldo.ianni@lngs.infn.it, osmirnov@jinr.ru}, A.~M.~Ianni\inst{4}, V.V.~Kobychev\inst{10},
G.~Korga\inst{1}, Y.~Kozlov\inst{8}, D.~Kryn\inst{6}, M.~Laubenstein\inst{1}, C.~Lendvai\inst{5},
M.~Leung\inst{4}, E.~Litvinovich\inst{8}, P.~Lombardi\inst{2}, I.~Machulin\inst{8}, 
D.~Manuzio\inst{11}, G.~Manuzio\inst{11},
F.~Masetti\inst{12}, U.~Mazzucato\inst{12}, K.~McCarty\inst{4}, E.~Meroni\inst{2},
L.~Miramonti\inst{2}, M.~E.~Monzani\inst{1}, V.~Muratova\inst{7},
L.~Niedermeier\inst{5}, L.~Oberauer\inst{5}, M.~Obolensky\inst{6},
F.~Ortica\inst{12}, M.~Pallavicini\inst{11}, L.~Papp\inst{2},
L.~Perasso\inst{2}, A.~Pocar\inst{4},\thanks{now at Stanford University},
R.~S.~Raghavan\inst{13}, G.~Ranucci\inst{2},
A.~Razeto\inst{1}, A.~Sabelnikov\inst{1}, C.~Salvo\inst{11},
S.~Schoenert\inst{9}, H.~Simgen\inst{9},
M.~Skorokhvatov\inst{8}, O.~Smirnov\inst{7},
A.~Sotnikov\inst{8}, S.~Sukhotin\inst{8}, Y.~Suvorov\inst{1}, V.~Tarasenkov\inst{8},
R.~Tartaglia\inst{1},
G.~Testera\inst{11}, D.~Vignaud\inst{6}, R. B.~Vogelaar\inst{13}, F.~Von~Feilitzsch\inst{5},
V.~Vyrodov\inst{8}, M.~W\'ojcik\inst{14}, O.~Zaimidoroga\inst{7} \and G.~Zuzel\inst{9}}

\institute{INFN Laboratori Nazionali del Gran Sasso, SS 17 bis Km 18+910, I-67010 Assergi(AQ), Italy \and
Dipartimento di Fisica Universit\`a and INFN, Milano, Via Celoria, 16 I-20133 Milano, Italy \and
Department of Chemical Engineering, A-217 Engineering Quadrangle, Princeton NJ 08544-5263, USA \and
Department of Physics, Princeton University, Jadwin Hall, Washington Rd, Princeton NJ 08544-0708, USA \and
Technische Universit\"at M\"unchen, James Franck Strasse, E15 D-85747, Garching, Germany \and
Astroparticule et Cosmologie APC, Coll\`ege de France, 11 place Marcelin Berthelot, 75231 Paris Cedex 05, France \and
Joint Institute for Nuclear Research, 141980 Dubna, Russia \and
RRC Kurchatov Institute, Kurchatov Sq.1, 123182 Moscow, Russia \and
Max-Planck-Institut fuer Kernphysik,Postfach 103 980 D-69029, Heidelberg, Germany \and
Kiev Institute for Nuclear Research, 29 Prospekt Nauki 06380 Kiev, Ukraine \and
Dipartimento di Fisica Universit\`a and INFN, Genova, Via Dodecaneso,33 I-16146 Genova, Italy \and
Dipartimento di Chimica Universit\`a, Perugia, Via Elce di Sotto, 8 I-06123, Perugia, Italy \and
Physics Department, Virginia Polytechnic Institute and State University, Robeson Hall, Blacksburg, VA
24061-0435, USA \and
M.Smoluchowski Institute of Physics, Jagellonian University, PL-30059 Krakow, Poland}


\abstract{
Electron antineutrino interactions above the inverse beta decay energy of protons 
($E_{\bar{\nu}_e}>$1.8~MeV) were looked 
for with the Borexino Counting Test Facility (CTF). 
One candidate event survived after rejection of background, which included muon-induced neutrons and random 
coincidences. An upper limit on the solar $\bar{\nu}_{e}$ flux, assumed having the $^8$B solar neutrino energy spectrum, 
of 1.1$\times$10$^{5}$~cm$^{-2}$~s$^{-1}$ (90\% C.L.) was set with a 7.8 ton $\times$ year exposure. 
This upper limit
corresponds to a solar neutrino transition probability, $\nu_{e} \rightarrow \bar{\nu}_{e}$,
of 0.02 (90\% C.L.). Predictions for antineutrino detection with Borexino, including geoneutrinos, 
are discussed on the basis of background measurements performed with the CTF.
\PACS{13.15.+g,14.60.St,13.40.Em,96.60.Hv,96.60.qd,23.40.Bw}
\keywords{neutrino magnetic moment, neutrino interactions, solar neutrinos, geoneutrinos, liquid scintillator detector}
}

\titlerunning{Electron antineutrinos at Gran Sasso with CTF and Borexino}
\authorrunning{Borexino collaboration, M.~Balata et al.}
\maketitle

\section{Introduction}

We report the results of the search for $\bar{\nu}_{e}$'s with the Counting Test Facility (CTF) 
for the Borexino experiment~\cite{ctf1,ctf1_2,Borexinodetector}. The CTF detector is located at the Gran Sasso underground laboratory, far away from
nuclear reactors, and thanks to its very low radioactive contamination, can detect antineutrinos from other sources with extremely low backgrounds. 
Known electron antineutrino sources include: (1) reactor $\bar{\nu}_{e}$'s, with
expected mean count in the CTF of 0.18~ev/y, and (2)
$\bar{\nu}_{e}$'s from the beta decays in chains of long-lived, natural radioactive isotopes (especially $^{238}$U and $^{232}$Th) distributed in the Earth interior (geoneutrinos).
Evidence of the latter was recently claimed by the KamLAND collaboration \cite{KamLAND_Geo}.  

A small antineutrino flux from the Sun is currently not completely excluded. One possible production mechanism
is neutrino-antineutrino
conversion due to spin-flavour precession (SFP), induced by a neutrino  
transition magnetic moment and originally proposed as a possible
solution to the observed solar neutrino deficit \cite{Schechter81,Akhmedov88,Lim88,Balantekin}%
\footnote{The model demands a non-vanishing neutrino magnetic moment at the
level of 10$^{-12}$-10$^{-11}$~$\mu_{B}$. An alternative model of
antineutrino production in $\nu$ decays in schemes with spontaneous
violation of lepton number was considered in \cite{Schechter82,Gelmini84,Gonzalez89,Beacom02}.%
}. This could be a sub-dominant process in addition to the MSW-LMA solution of the solar neutrino problem%
\footnote{A discussion on the robustness of the MSW-LMA solution is presented in
\cite{RobustLMA}.%
}. The interest in searching for a large neutrino magnetic moment was recently revived,
mainly because of the new experimental data available from  KamLAND  
and in view of forthcoming low energy solar neutrino detectors such as Borexino. 
A discussion of the constraints on the possible Majorana neutrino
transition magnetic moment from existing and near future experiments
can be found in~\cite{Joshipura,KamLAND_Chauhan,Kang,Rashba-1,Rashba-2}.
In particular, it was shown that a random
magnetic field in the convection zone of the Sun can enhance the rate
of $\bar{\nu}$'s through spin/flavour conversion~\cite{Rashba-1}. Such
enhancement would improve the detectability of a neutrino magnetic
moment down to the level of $10^{-12}\mu_{B}$. The CTF detector itself
demonstrated a sensitivity to the solar neutrino magnetic moment of $5.5\times10^{-10}\mu_{B}$~\cite{EMCTF2}. 

In this paper we mainly set a limit on the solar antineutrino
flux. We also discuss the sensitivity of CTF to geoneutrinos, as well
as the discovery potential of the Borexino experiment. 

\section{Experimental Method and advantages of CTF}

The inverse-beta decay of protons

\begin{equation}
\overline{\nu}_{e}+p\longrightarrow e^{+}+n,\label{eq1}
\end{equation}
with a threshold of 1.806~MeV, is the dominant electron-antineutrino interaction in 
liquid scintillator (LS) or water. The cross section
for this process is two orders of magnitude higher than that for
($\bar{\nu}_{e}$,e) elastic scattering. In organic scintillators 
this reaction generates a prompt signal from the positron and a
delayed one, following the neutron capture on protons

\begin{equation}
n+p\longrightarrow d+\gamma\;(2.22MeV).
\end{equation}
The total energy released by the positron after annihilation is $E=T+2m_{e}c^{2}$,
where $T$ is the positron kinetic energy. Neglecting the small neutron
recoil, the {\emph{visible energy}} can be written as $E_{\bar{\nu}_{e}}-0.78$~MeV.
The capture of thermalized neutrons on protons with a mean life-time
of $\sim 200\div250$ $\mu s$ provides a tag for this reaction in
a LS detector, allowing significant reduction of background. Neutron captures on $^{12}$C is also possible but
with a much smaller probability. 

In existing water Cherenkov detectors the delayed $2.22$ MeV $\gamma $ is below the
detection threshold and hence a  positron from inverse-beta decay is indistinguishable
from an electron or a $\gamma$,  
making such detectors significantly less sensitive than LS detectors.
In fact, the recent Super-Kamiokande (22 kton water Cherenkov detector) limit
for solar antineutrino flux $\phi_{\bar{\nu}_{e}}<1.32\times10^{4}$~cm$^{2}$~s$^{-1}$
in the energy region $8<E_{\bar{\nu}}<20$~MeV (90\% C.L.) \cite{SuperK03}
was significantly improved by KamLAND (1 kton LS detector), 
$\phi_{\bar{\nu}_{e}}<3.7\times10^{2}$~cm$^{2}$~s$^{-1}$~\cite{KamLAND04}
 (90\% C.L.) in the energy region $8.3 \leq E_{\bar{\nu}_e} \leq 14.8$ MeV.
The current experimental constraints on the solar antineutrinos flux are listed in Tab.~\ref{cap:BestLimits}.
The best limit is obtained for energies above 8.3 MeV.
The region below 4.0 MeV has not been explored. The CTF detector provides a unique possibility
to look for evidence of a solar antineutrino flux at low energy. 
The CTF can detect $\bar{\nu}_e$'s at the inverse-beta decay threshold with very little background from nuclear 
reactors and from cosmogenic radioactivity (approximately 7 times lower than at Kamioka).

\begin{table*}
\caption{Experimental constraints
on the flux of solar $\bar{{\nu}_e}$'s. $\phi_{\bar{\nu}_e}^{meas}$
is the limit on the flux within the experimental energy range 
(90\% C.L.) $\phi_{\bar{\nu}_e}^{tot}$ is the limit scaled to the
total energy range. BP04~\cite{BP04} gives a $\bar{\nu}_{e}$'s
flux from $^{8}$B equal to (5.79$\pm$1.33)$\times10^{6}$ cm$^{-2}$s$^{-1}$.
Here SK stands for SuperKamiokande and KL for KamLAND.}\label{cap:BestLimits}

\begin{tabular}{lccccc}
\hline\
 & LSD & SK & KL & SNO & CTF\\
\hline\hline
 Exposure & 0.094 & 92.2 & 0.28 & 0.584 & 0.0078\\
 kt$\times$yr &   &      &      &       &      \\
\hline
$\phi^{meas.}_{\bar{\nu}_e}$ & $<0.46\times$10$^5$ &  $<1.32\times$10$^4$ & $<3.7\times$10$^2$ & 
$<3.4\times$10$^4$ & $<1.06\times$10$^5$\\
cm$^{-2}$s$^{-1}$  &   &   &   &   &\\
\hline
$\phi^{tot}_{\bar{\nu}_e}$ & $<1\times$10$^5$ &  $<4\times$10$^4$ & $<1.3\times$10$^3$ & 
$<5.2\times$10$^4$ & $<1.08\times$10$^5$\\
cm$^{-2}$s$^{-1}$  &   &   &   &   &\\
\hline
$\frac{\phi_{\bar{\nu}_e}}{\phi_{\bar{\nu}_e}(^8B)}$ & $\leq 1.7\times$10$^{-2}$ &  
$\leq0.7\times$10$^{-2}$ & $\leq2.2\times$10$^{-4}$ & 
$\leq 1\times$10$^{-2}$ & $\leq 1.9\times$10$^{-2}$\\
\hline
$E_{\bar{\nu}_e}$ range & [7,17] & [8,20] & [8.3,14.8] & [4,14.8] & [1.8,20]\\
MeV               &        &        &            &          &         \\
\hline
Reference,&\cite{LSD}&\cite{SuperK03}&\cite{KamLAND04}&\cite{SNO}&this paper\\
year&1996&2003&2004&2004& \\
\hline
\end{tabular}
\end{table*}

\section{The CTF detector}

CTF is an unsegmented liquid scintillator detector. Its active volume, a large amount of 
liquid scintillator contained in a transparent spherical nylon shell, 2 m diameter and 0.5 mm thick, is immersed in 1000 m$^3$ of high purity shielding water. 100 PMTs, mounted on an open structure immersed in the water, surround the nylon sphere and detect the light from events in the scintillator. 
The water, contained in a cylindrical tank (10 m diameter, 11 m high), shields the scintillator against $\gamma$ radiation emitted by radioactive contaminants
in the PMTs and their support structure as well as against $\gamma$'s following the capture of 
neutrons generated within the walls of the experimental hall.
Another 16 upward-looking PMTs of an active muon veto system (MVS) are mounted on the bottom of the tank. They detect the Cherenkov light of muons that cross the water without intersecting the scintillator. 
The muon-veto was tuned to maximize the muon tagging efficiency while minimizing the probability of scintillation light pickup for
sub MeV events (CTF was optimized to study backgrounds in the  {[}0.25,0.8{]}~MeV energy range, where Borexino will look for $^7$Be solar neutrino interactions~\cite{Borexinodetector}).
A more detailed description of the CTF detector can be found in~\cite{ctf1,ctf1_2}. 

The CTF has been in operation since 1993.  
During the 1993-1995 campaign (CTF1), the detector was filled with $\sim$ 4 tons of 
pseudocumene (PC, 1,2,4-trimethylbenzene, C$_{6}$H$_{3}$(CH$_{3}$)$_{3}$, $\rho=0.88 $ g/cm$^{3}$) to which PPO 
(2,5- Diphenyloxazole, \,\,\, C$_{15}$H$_{11}$NO) was added as a wavelength shifter in low concentration (1.5 g/l). 
This run was focused on studying backgrounds for the Borexino scintillator~\cite{ctf1}. 
In 1999, CTF was run again (CTF2), this time with PXE (1-Phenyl-1-xylylethane, C$_{16}$H$_{18}$, $\rho=0.995$
g/cm$^{3}$) scintillator. It was upgraded to include an active muon-veto; also, a second, 125 micron thick, nylon
membrane was added in the water space between the PMTs and the scintillator, 
aiming to suppress Rn diffusion from the periphery to the center of the detector~\cite{CTF2paper}. These two additions turned
CTF into a sensitive detector in the field of rare events physics, as proven by the results
in~\cite{EMCTF2,EDCTF2,Paulipaper,B8decay,NucleonDecay,RDCTF2,Review}.
In 2002 a third campaign with PC+PPO liquid scintillator began (CTF3); it is still in progress to finalize the purification strategy for the Borexino scintillator.

The electronics of CTF are designed to record fast delayed coincidences
without appreciable dead time. Time and charge information of the
PMT pulses of an event are recorded by a set of ADCs and TDCs (group 1 chain). 
During the acquisition time, a second set of
ADCs and TDCs (group 2 chain) is sensitive to a possible other event occurring
within 8.3 ms. The coincidence time between the two chains is measured
by means of a long range TDC. Subsequent events are ignored until the
group 1 chain is ready again. The group 1 trigger is fired when 6 PMT hits occur within a 30 ns from each other.
The corresponding energy threshold is measured to be $\sim$20~keV at 50\%
detection efficiency; 99\% detection efficiency corresponds to an
energy threshold of 90~keV. The group 2 chain trigger is set at $\sim$150~keV. 
To avoid retriggers due to PMT after-pulses and 
cosmogenic short-lived isotopes, the group 2 chain is vetoed for 20 $\mu$s
after each MVS trigger; this time region is excluded from the analysis. The energy response of the detector
is calibrated run-by-run using the light yield obtained by fitting the $^{14}$C energy spectrum: on average $\sim$3.8 
photoelectrons (p.e.) per PMT are detected for 1 MeV recoiling electron at a random position within the
detector volume. Electronics of each channel from the PMT to the ADC is linear up to 20 p.e., which guarantees a linear energy response for events below 4.5 MeV. An independent
chain of electronics with flash ADCs was also used in CTF2 and CTF3 in order to increase the dynamic range of the detector.
The shape of the total signal of the detector
(analog sum of all 100 PMTs channels) is digitized by an 8 bit Transient
Time Recorder (TTR) for 1 $\mu$s with 5 ns resolution.

In the present study we use CTF3 data collected during 855.6 days of data taking
(764.2 days of live-time) to search for $\bar{\nu}_{e}$'s interactions. 
Previous analyses \cite{EMCTF2,EDCTF2,RDCTF2,Review} selected events from only the innermost part of the scintillator
 in order to improve the specific signal-to-noise performance.
Since inverse-beta decay has an easily recognizable signature
(the coincidence between the positron and the delayed $\gamma$-ray following neutron capture), 
the whole detector volume has been used for this study; this resulted in no noticeable random background.

\section{Data selection and backgrounds}

\begin{figure}
\resizebox{0.5\textwidth}{!}{%
  \includegraphics{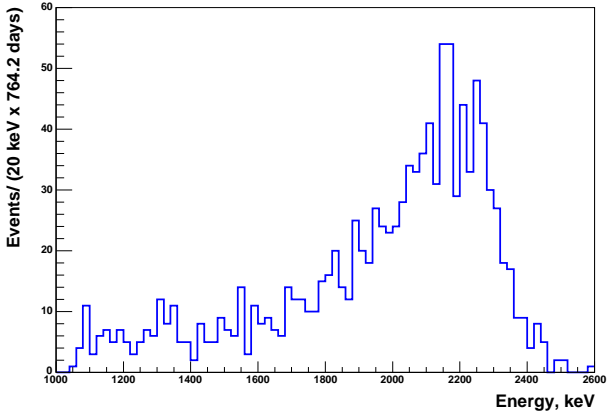}
}
\caption{Energy spectrum registered by the second
group of the electronics for neutron capture candidates 
(in coincidence with muon-tagged events in the first group).
 The full absorption peak of 2.22~MeV gamma's emitted
in the muon-induced neutron capture on proton is clearly seen at $\sim$ 2.2~MeV (the scale is calibrated with electrons, the
position of the gamma is shifted toward lower energies due to the ionization quenching effect).}
\label{Figure:EnergyGamma}       
\end{figure}

\begin{figure}
\resizebox{0.5\textwidth}{!}{%
  \includegraphics{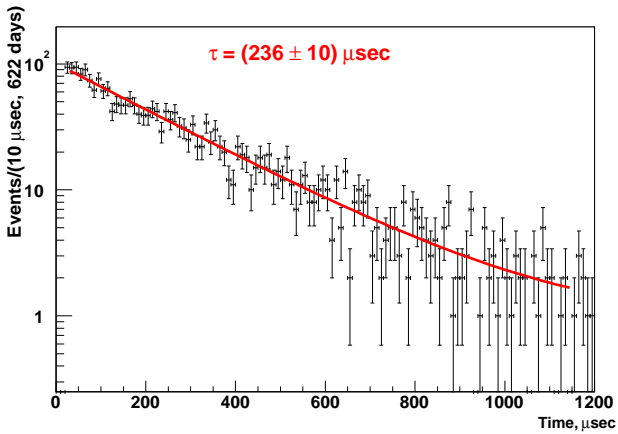}
}
\caption{Coincidence time between muon and 2.22 MeV
gamma produced in neutron capture on proton. The fit has been performed using an exponential plus a constant function.}
\label{Figure:TimeGamma}       
\end{figure}

Candidate events were searched among all correlated (in space and time) events occurring within
2~ms one after another, excluding coincidence times smaller than 20~$\mu$s.
The energy of the prompt event was set to be 0.85~MeV$<E<$20~MeV. The lower limit is defined by the
threshold of the inverse-beta decay reaction (visible energy of 1.02~MeV) taking into account the finite energy resolution of the
detector, $\sigma(E)$(MeV)$\sim 0.1\sqrt{E/1MeV}$. The energy of the second event was required to be  1.1~MeV$<E<$2.6~MeV for detecting 
the 2.2 MeV $\gamma$-ray with high efficiency and avoiding the delayed
$^{214}$Bi-$^{214}$Po coincidences.
The energy calibration of the first group of the electronics was performed
using $^{14}$C events and checked at higher energies using the first event of the delayed
$^{214}$Bi-$^{214}$Po coincidences (originating from $^{222}$Rn in the LS); $^{214}$Bi $\beta$ decays with $Q$ value of 3.2 MeV.
The energy and spatial resolution
of the CTF3 detector are very close to those of CTF1~\cite{ctf1,Resol}. The energy calibration of the second group of the electronics was 
checked using the 2.22 gamma-ray from neutron
capture on protons, which is a prominent feature of the group 2 energy spectrum
(see Fig.~\ref{Figure:EnergyGamma}).
Coincidence times between the first and second events are shown in
Fig.~\ref{Figure:TimeGamma}. The measured life-time of 236 $\mu$s lies, not surprisingly,
between the simulated values for neutron capture in water (220 $\mu$s) and PC (250 $\mu$s); 
indeed, a fraction of the detected captures happen in the shielding water.

\begin{figure}
\resizebox{0.5\textwidth}{!}{%
  \includegraphics{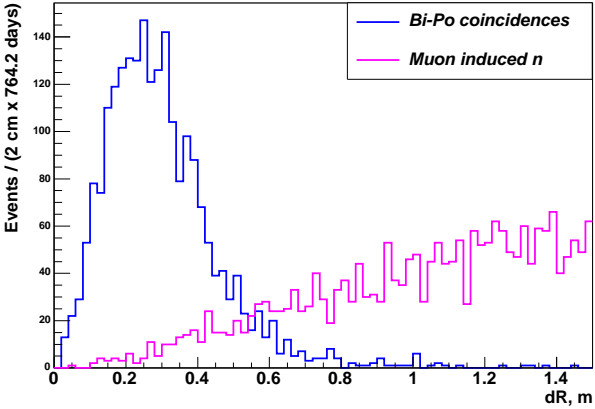}
}
\caption{Reconstructed distance between the first and the
second event for the $^{214}$Bi-$^{214}$Po coincidence events and
for muon-induced neutron events. In the last case the reconstructed distance cannot
be assigned to a real distance and should be treated as a convenient
parameter for the muon induced/scintillation events discrimination.}
\label{Figure:dR}       
\end{figure}

The position resolution of the detector can be measured using delayed coincidences, and is $\sim$10 cm (1$\sigma$) 
for $^{214}$Bi-$^{214}$Po events. 
In the case of muon tracks, the reconstruction code
gives a point-like weighted position of the event, which often falls outside the detector's
active volume. Such feature is a useful tool
for muon event discrimination. The reconstructed distance, dR, between the
first and second event of $^{214}$Bi-$^{214}$Po coincidences
and of muon-induced neutron events is shown in Fig.\ref{Figure:dR}. 
In the latter case, muons that "skim" the scintillator volume can generate a prompt signal falling in the group 1 
energy cut and produce a neutron which is then captured, giving rise to a coincidence event.
A cut on the distance between the two events in coincidence of $dR<$0.7 m, optimized using simulated events, was chosen 
for the antineutrino event selection; this cut preserves 80\% of the sought for antineutrino induced scintillation events.

%

\begin{figure}
\resizebox{0.5\textwidth}{!}{%
  \includegraphics{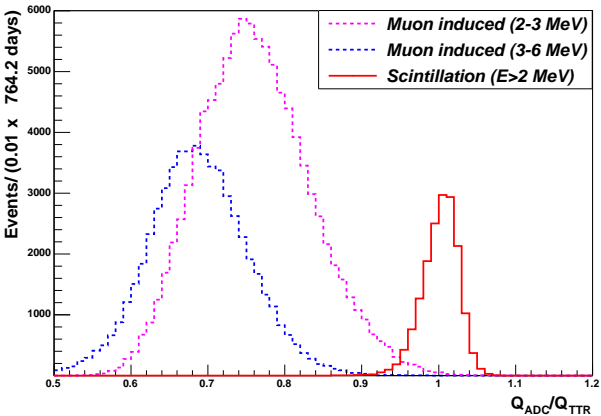}
}
\caption{Efficiency of the muon/scintillation events
discrimination in 2 MeV - 6.0 MeV energy window using $r=$Q$_{ADC}$/Q$_{TTR}$. 
Scintillation events are integrated by the ADC and TTR in the same
way, providing $r$ greater than 0.9, while for muon events the fraction
of charge collected by ADC is less than that integrated with TTR.}
\label{Figure:QAdcVsQTdc}       
\end{figure}

\begin{figure}
\resizebox{0.5\textwidth}{!}{%
  \includegraphics{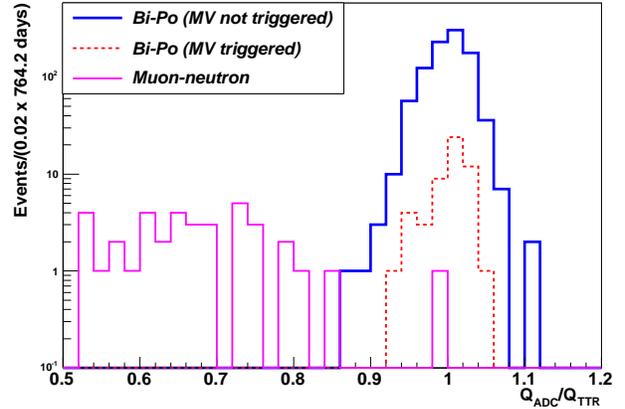}
}
\caption{ $r=$Q$_{ADC}$/Q$_{TTR}$ ratio
for 3 taggable event classes (at E$>$2~MeV). One can see that the $r$ value for
the scintillation events from $^{214}$Bi-$^{214}$Po coincidences
are around $r=1.0$ independently of the Muon Veto System trigger,
while for the muon events, followed by correlated neutron, the mean value of $r$ is much
lower. }
\label{Figure:QAdcVsQTdc_Radon}       
\end{figure}

\begin{figure}
\resizebox{0.5\textwidth}{!}{%
  \includegraphics{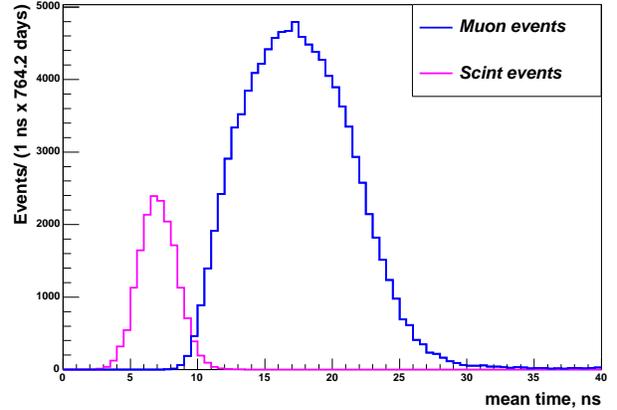}
}
\caption{Mean arrival time of the light signals for scintillation
and muon events (E$>$2~MeV).}
\label{MeanTime}       
\end{figure}

\begin{figure}
\resizebox{0.5\textwidth}{!}{%
  \includegraphics{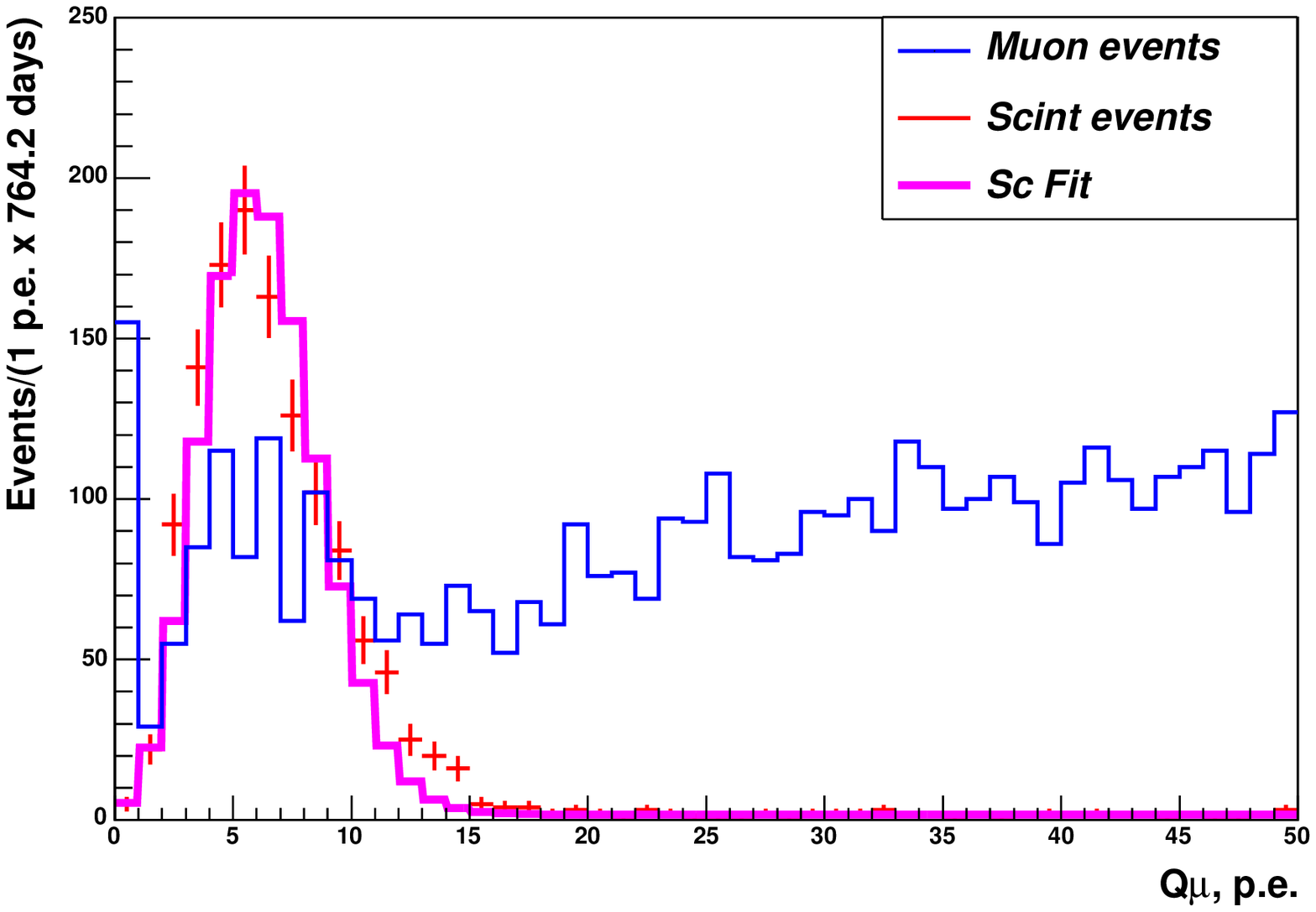}
}
\caption{Charge collected on the MVS for events identified
as muons and scintillation ones, respectively (2.0-6.0 MeV energy
window). The scintillation light pickup at the MVS is at the level
of 2~p.e. for 1~MeV energy deposit in the active detector and can be modeled with 
a Poisson-like distribution (shown with a thick line). }
\label{Figure:QMu}       
\end{figure}

In the present analysis, we use 
the MVS tag only for events with
E$<$2.0~MeV, where the probability of scintillation event tagging as a muon
is less than 1\%. In order to minimize the probability of discarding good candidate prompt 
events above 2 MeV by mistakenly tagging them as muons (using the MVS triggered by the large scintillation light produced), 
the muon identification was performed using specific
features of muon and scintillation events in the energy interval 2.0-6.0~MeV. 
The following three criteria were used for scintillation/muon events discrimination:

\begin{enumerate}
\item \textbf{ratio of the charge measured by the ADCs of the main system
to the charge measured by TTR, $r=Q_{ADC}/Q_{TTR}$.} The main trigger (i.e. that of the 100 PMTs looking at the scintillator)
is activated either when 6 photomultipliers fire within a 30 ns window (the threshold for each PMT is set at the level of 0.2
p.e.) or when 4 photomultipliers of the MVS are above threshold (set at 1.5 p.e.). The timing of the
main system ADCs gate (with 100 PMTs) hence depends on where the event is (water or scintillator):
Cherenkov photons precede scintillation pulses by 3-4 ns. 
For muon-induced events the gate
of the ADCs arrives with a few nanoseconds delay and thus part of the signal
is not integrated.
The ratio of the total charge, $Q_{ADC}$, measured with the ADCs of the main
system to the total charge estimated integrating the TTR signal, $Q_{TTR}$,
provides a good tag for muon/scintillation events discrimination in
2.0~MeV - 6.0~MeV energy window. Fig. \ref{Figure:QAdcVsQTdc} and Fig. \ref{Figure:QAdcVsQTdc_Radon}
illustrate the efficiency of the method. Above 6.0~MeV,
the ADCs of the main system saturate and this method is not directly applicable.
\item \textbf{mean arrival time of light registered by the system of 100 PMTs, $\overline{t}$}.
For the scintillation events the mean time $\overline{t}$ is lower
than for the muon induced events (see Fig.\ref{MeanTime}), as explained in~\cite{ctf1}. We used $\overline{t}<$12
ns as the scintillation acceptance criteria. 
This cut preserves the maximum number of scintillation
events (99.8\% at E$>$2~MeV) and rejects about 95\% of muons.
\item \textbf{the amount of light seen by the MVS, $Q_{\mu}$}. Fig. \ref{Figure:QMu}
illustrates the discrimination procedure. The scintillation light
pickup for the MVS system is 2~p.e. for 1 MeV energy
deposit in the active detector. In the energy range 2.0-6.0 MeV, $Q_{\mu}<$30
p.e. has been used as scintillation acceptance criteria; at higher energies the upper limit has been
set at $Q_{\mu}<$100~p.e. (which allows to separate a 20~MeV energy
deposit in the main detector seen by the MVS at the level of 5$\sigma$). 
\end{enumerate}
The analysis of the candidate events energy based on the calibration
with $Q_{TTR}$ (instead of $Q_{ADC}$ used at sub-MeV energies) showed
that the reconstructed energy of all but one event falls out of the
window of interest for the solar antineutrino analysis (0.85-20 MeV). The details are presented in Tab.~\ref{CandidatesSelection}.

\begin{table*}
\caption{Candidates selection. The initial selection
was performed on E$_{2}$ and dT (1.1$<E_{2}<$2.6 MeV; 20 $\mu$s$<dT<$2
ms). The MVS hardware tag was not used at E$_{1}>2$ MeV.}
\label{CandidatesSelection}
\begin{tabular}{lcccc}
\hline\noalign{\smallskip}
Cut&
\multicolumn{3}{c}{Candidate events in corresponding E$_{1}$ (MeV) range }& Total\\ 
 & $0.85<E_{1}<2.0$ & $2.0<E_{1}<6.0$ & $6.0<E_{1}$ &  \\
\hline 
Total& 27& 130& 956& 1113\\
\hline 
$dR<70$ cm & 2& 46& 195& 243\\
\hline 
MVS tag& 2& 0& 0& 2\\
\hline 
Q$_{ADC}$/Q$_{TTR}>$0.9& --& 5& --& --\\
\hline 
Q$_{\mu}<Q_{lim}$& 6 (Q$_{lim}$=30)&  6 (Q$_{lim}$=30)& 62 (Q$_{lim}$=100)& 74\\
\hline
$\overline{t}<12$ ns& 4& 39& 146& 189\\
\hline
all cuts& 0& 1& 5& 6\\
\hline
E$_{Rec_{1}}<20$ MeV& 0& 1& 0&1\\
\hline
\end{tabular}
\end{table*}

Physical background signals for the antineutrino analysis are coming
mainly from reactors $\bar{\nu}_{e}$'s. We have estimated this
background source considering 42 nuclear reactors in Europe~\cite{rotunno}
and using the best fit estimation for the oscillation parameters~\cite{KamLAND2005}
and $\bar{\nu}_{e}$'s spectra from~\cite{HuberSchwetz}. The contribution
of the geoneutrinos is negligible (see Section 6).

Other sources of background are listed below in the order of their relative importance.

(1) \textbf{neutrons produced by cosmic muons}. The residual cosmic
muon flux at the Gran Sasso depth (3800 mwe) has a rate of 1.2 count/$m^{2}$/h
and an average energy of $E$ = 320 GeV \cite{MACRO}. Cosmic muons
are identified with high efficiency by the muon tagging described
above if they pass through the detector. On the contrary, neutrons
produced by muons outside the detector can produce a fake event for
the antineutrino search. In particular, a high energy neutron produced
in the surrounding rocks may enter the detector and scatter off a
proton (or excite low-lying levels of $^{12}$C). In this case
the proton (or gamma quantum) gives the prompt signal and the recoiled
neutron, once thermalized and captured, produces the delayed event. There is no
special tag for these events, the probability of this background was
evaluated by Monte Carlo method using the FLUKA code~\cite{FLUKA,C11paper}. We have not considered neutrons produced by spontaneous fissions or ($\alpha$,n) reactions in
the rocks of the undeground laboratory since they constitute a smaller flux at E$>$10~MeV than that of neutrons induced by muons. 
This can be easily seen by comparing the flux determined by using the neutron yield per muon~\cite{MuonNeutronProduction} 
against the predicted flux induced by radioactivity~\cite{NeutronFluxRadioactivity}.

(2) \textbf{accidental coincidences}. Their probability was estimated
using selected events falling in an off-time delayed window, 2-8 ms,
after the prompt event. The same energy cuts as in the antineutrino analysis were applied to select random coincidences events.

(3) \textbf{cosmogenic radioactivity}. In organic scintillator a possible
residual cosmogenic background may originate from muons crossing the
scintillator. As discussed in~\cite{Hagner} a certain number of
radioactive isotopes can be produced on $^{12}$C nuclei in the
CTF scintillator. Among the possible isotopes contributing to backgrounds $^{8}$He and $^{9}$Li
are of particular interest for the search of antineutrinos: 
 $^8$He can decay in $\beta^{-}n$ with $t_{1/2}=0.12$~s
(Q=10.7~MeV, 16\%); the $^9$Li can decay in $\beta^- n 2 \alpha$ with $t_{1/2}=0.18$~s (Q=13.6~MeV, 45.5\%).
We have searched for such events after each tagged
muon. 
In particular, in order to reduce this background
we have checked the arrival time of the muon preceding every candidate event.
Muons crossing the LS produce a very large signal in CTF and can be easily discriminated. A 2~s time window after such
events was excluded from the analysis. The cosmogenic
background is thus reduced to 10$^{-4}$ events for all
the period of the data taking.

(4) \textbf{$^{13}$C}. As it has been discussed in~\cite{KamLAND_Geo} a high 
contamination of $^{210}$Po in the LS can be a source of fake events in $\bar{\nu}_e$'s searches. 
In fact, the $\alpha$ decay of $^{210}$Po can induce the reaction $^{13}$C($\alpha$,n)$^{16}$O which produces a neutron. This
sequence is a source of a correlated background because the produced neutron can first scatter off a proton which gives a
prompt signal and, later be captured; another possibility is that the prompt is produced by the de-excitation of
$^{12}$C after $^{12}$C(n,n$\gamma$)$^{12}$C ($E_\gamma$=4.4~MeV) or the de-excitation of $^{16}$O. In
KamLAND~\cite{KamLAND_Geo,NANP05} the background induced by $^{13}$C is
estimated to be 42$\pm$11 events with a measured activity of $^{210}$Po on the order of 22~Bq and an exposure of
~5$\times$10$^{31}$ protons$\times$year. In the CTF the $^{210}$Po activity is measured to be $\sim$20~$\mu$Bq/ton 
($\sim$ 10$^3$ times lower than that of KamLAND) and
this background is therefore negligible (1 ton of CTF3 scintillator contains $\sim 6 \times 10^{28}$ protons).

A summary of the background and systematic uncertainties of the 7.8 ton$\times$yr
exposure for the search of $\bar{\nu}_e$'s from the Sun is reported in Tab.~\ref{cap:Errors}.

\begin{table}
\caption{Estimated backgrounds and systematic uncertainties
for 764.2 days of CTF livetime, equivalent to 7.8 ton $\times$ year exposure
 (62\% efficiency taken into account).}
\label{cap:Errors}
\begin{tabular}{ll}
\hline
\bf{Background} & \bf{Expected events} \\ \hline
accidental coincidences & 0.08 \\
reactor antineutrinos & $0.37$ \\
fast n,p scattering&$0.8\pm0.3$\\
fast n on $^{12}$C (4.4 MeV)&$0.07\pm0.03$\\ \hline
\bf{Systematic uncertainties }& \% \\ \hline
efficiency, $\epsilon$ & 2 \\
number of protons, $N_{p}$ & 3.4\\
Energy threshold&$<2$\\
Livetime & 2 \\
\hline
\end{tabular}\end{table}

\section{Analysis}

In the Monte Carlo simulation of the detector efficiency
events were generated in accordance with the $^{8}$B
solar neutrino spectrum inside the inner vessel and in an adjacent
water layer of 50 cm. The gamma and electron/positron showers were followed
using the EGS-4 code \cite{EGS4}. Neutron diffusion was also taken
into account. The detector energy and spatial resolution 
was calibrated with radioactive sources and modeled via MC method.
The total detection efficiency found after applying all cuts described above is 62$\pm$2\%
(see Tab.~\ref{CutsEff} for the details). 

\begin{table*}
\caption{Individual cut efficiencies (only scintillation events
acceptance efficiencies are shown). All cuts were tuned to have maximum
acceptance efficiency for the scintillation events.}
\label{CutsEff}
\begin{tabular}{lc}
\hline\noalign{\smallskip} 
Cut& Scintillation event \\
   & acceptance efficiency \\
\hline
$dR<70$ cm (CTF detector, MC, including n/$\gamma$ escape)& 79$\pm$1\\
($dR<70$ cm in an infinite media, MC) & (99.8) \\
\hline
20 $\mu$s$<dT<$2 ms&92.5$\pm$0.1\\ \hline
0.85 MeV$<$E$_{1}<$20 MeV&99.2$\pm$0.2\\ \hline
1.1 MeV$<$E$_{2}<$2.6 MeV&88$\pm$1\\ \hline
muons discrimination in E$<$2 MeV (MVS)&
$>$98\\ \hline
muons discrimination in 2$<$E$<$6.0 MeV (r$>$0.9 and Q$_{\mu}<$30)&
$>$99\\ \hline
muons discrimination in $6.0<E<20$ MeV ( Q$_{\mu}<100$ and $\overline{t}<12$
ns)&$>99$\\ \hline
total&62$\pm$2\\
\hline
\end{tabular}\end{table*}

As noted before (see Section 4) only one candidate event was found. The event's characteristics  
are reported in Tab.~\ref{Table:Candidate},
where Q$_{ADC}$/Q$_{TTR}$ is the muons discrimination
variable described above, $dt$ is the coincidence time, $R$ is the
reconstructed event vertex position, $dR$ is the reconstructed distance
between the prompt and delayed events, t$_{\mu}$ is the time passed from the moment of registering the previous muon (used to discriminate background from the short-living
cosmogenic isotopes), $\overline{t}$ is the mean arrival time of the signals detected
by PMTs and, Q$_{\mu}$ is the charge collected by ADCs of the MVS. The
candidate event was tagged by the hardware muon-veto. This fact could
be due to the scintillation light pickup by the muon-veto in the case of scintillation event, as well
as due to the Cherenkov light produced by a muon. According
to the analysis criteria presented above, this event has all the characteristics of a
scintillation event. 

\begin{table*}
\caption{Main features of the candidate event. The
prompt event was tagged by the hardware muon veto but the amount of detected light
was just at the muon veto system trigger level. Both prompt and delayed events are reconstructed
close to the detector center (with a weighted position of about 40 cm away from it). See text for further details.}
\label{Table:Candidate}
\begin{tabular}{lcccccccc}
\hline
 & E   & Q$_{ADC}$/Q$_{TTR}$& dR &  dt, $\mu$s& R & t$_{\mu}$ & Q$_{\mu}$& $\overline{t}$\\
 & MeV &                    & cm &  $\mu$s    & cm & s        &          & ns            \\
\hline
\hline 
prompt& $4.37\pm0.44$& 1.04& --& --& 30& & 7.0& 3.7\\
\hline 
delayed& $2.23\pm0.13$& --&--&--&56& & --& 5.4\\
\hline 
 & & & 28& 207& 42& 46.9& &
\\
\hline
\end{tabular}\end{table*}

We note that the prompt energy of the candidate event is 4.37~MeV,
which coincides, within experimental errors, with the energy of the first excited level of $^{12}$C
of 4.4~MeV. This, together with
the fact that the muon veto was triggered, could be due to the excitation
of the first $^{12}$C level by a fast neutron produced by a muon
passing outside the detector, near the water tank inner wall. In total we observed
20 events of 4.4$\pm$0.6 MeV energy in coincidence with a 2.22 MeV
neutron capture gamma (1.8 MeV$<$E$<$2.6 MeV), all but one (the antineutrino candidate)
identified as muons during the analysis and tagged by the MVS. The
probability of this type of events for muons passing close to the
detector walls (i.e. escaping identification by the muon veto system
and by the \char`\"{}muon cuts\char`\"{}) was estimated by MC method, and found to be at the level of fraction of an event
for the time period of interest (see Tab.~\ref{cap:Errors}).
Although we cannot completely exclude that the selected event was caused
by a passing muon, it will be treated as an antineutrino candidate
event in the following analysis of the antineutrino flux limits. 

The hypothetical flux of $\bar{\nu}_{e}$'s
from $^{8}$B, assuming no spectral distortion, can be obtained from
the following equation:

\begin{equation}
\phi_{\bar{\nu}_{e}} = \frac{N_{\bar{\nu}_e}}{N_{p}\times t\times\epsilon\times\langle\sigma\rangle},
\label{eq2}
\end{equation}
where $N_{\bar{\nu}_{e}}$ is the number of detected events, $N_{p}=2.25\times10^{29}$ is the number of target protons,
$t=6.60\times10^{7}$ s is the live-time,
$\epsilon$=62\% is the mean detection efficiency, and $\langle\sigma\rangle=3.4\times10^{-42}$~cm$^{2}$ is 
the cross-section folded over the $^{8}$B spectrum in the energy
range of interest. An upper limit for the electron antineutrino flux, assuming no distortion in the $^{8}B$ spectrum, is derived below in light of the observed one
candidate event. A Bayesian approach was used with a constant prior
and a likelihood function defined as:

$$
L(s,b,\sigma_b,\sigma_\eta,n)=\int db^{\prime} \int d\eta Pois(\eta\cdot s+b^{\prime},n)\times Gaus(b-b^{\prime},\sigma_{b})
$$
\begin{equation}
\times Gaus(1-\eta,\sigma_\eta),\label{eq3}\end{equation}
where $Pois(\eta\cdot s+b,n)$ is a Poisson distribution with mean
value equal to $\eta\cdot s+b$, ($s$ is the expected signal and
$b$ is the background with uncertainty $\sigma_{b}$), and $n$
is the number of observed events ($n=1$ in this case); $\sigma_\eta$ is the total systematic
uncertainty and $Gaus(x_0-x,\sigma_{x})$ is a Gaussian with mean value
$x_0$ and standard deviation equal to $\sigma_{x}$.

An upper limit for the solar $\bar{\nu}_e$ flux of $\phi_{\bar{\nu}_e} <1.1\times10^{5}$~cm$^{-2}$~s$^{-1}$
is obtained from Eq.(\ref{eq2}) using data from Tab.~\ref{cap:Errors}
(for 1 candidate and 1.3$\pm$0.7 background events we expect 3.3
coincidences at 90\% C.L.). Using the BP04~\cite{BP04} standard
$^{8}$B flux for solar neutrinos we derive an upper limit for
the ratio of the antineutrino to neutrino fluxes of $\phi_{\bar{\nu}_{e}}/\phi_{^{8}B} < 1.9\times10^{-2}$
at 90\% C.L.

CTF owes its sensitivity to both the excellent radiopurity and
low reactor antineutrino background. Borexino
can open an interesting opportunity in searching for electron antineutrinos
from the Sun. The expected sensitivity is $\phi_{\bar{\nu}_{e}}/\phi_{^{8}B} \sim 1\times$10$^{-5}$
in 5 years. Such search is important for looking for a neutrino magnetic
moment and for studying the magnetic field inside the Sun~\cite{Friedland}.

\section{Antineutrinos from the Earth. Estimation of the Borexino discovery
potential based on the CTF results}

\begin{figure}\
\resizebox{0.5\textwidth}{!}{%
  \includegraphics{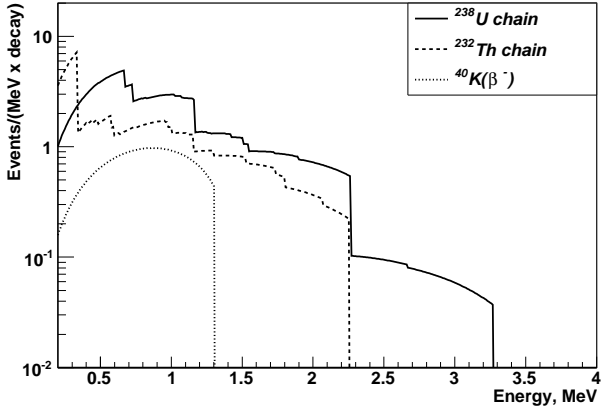}
}
\caption{The geoneutrino spectrum from the U, Th and K. Only antineutrinos from the U and Th 
decay chains have energies above the inverse-beta decay reaction threshold (1.8~MeV).}
\label{Figure:AntiNu}       
\end{figure}

In this section we discuss 
some features of a future measurement in Borexino of antineutrinos
generated in the Earth interior, on the basis of the data presented above. 
It is believed that about 40\% or more of the heat radiated
by the Earth has radiogenic origin~\cite{Fiorentini,Rothschild,Raghavan,Fogli,Nunokawa,Fiorentini050891}. 
The heat generated by radioactive decays of $^{238}$U, $^{232}$Th, their daughters 
and of $^{40}$K in the Earth is estimated at $\sim$30 TW
using a model based on the studies of composition of chondritic meteorites; 
a value of $\sim$20 TW is predicted by the so-called Bulk Silicate Earth model~\cite{Nunokawa,BSEarth1}. 
The heat produced by all the decays in the $^{238}$U chain is 9.5$\times$10$^{-5}$~W/kg, 
while 2.6$\times$10$^{-5}$~W/kg are generated by the decays in the $^{232}$Th chain. 
Six and four antineutrinos are emitted per full U and Th decay chain, respectively; the specific antineutrino
intensity is 7.46$\times$10$^{7}$~Bq/kg for U and 1.62$\times$10$^{7}$~Bq/kg for Th.
Current estimations of $^{238}$U abundancy (0.4$\times$10$^{17}$~kg) 
suggest that the crust alone should radiate
$\sim$3$\times$10$^{24}$ $\bar{\nu_{e}}$ from this source, 
corresponding to a flux of $\sim 10^{6}$~cm$^{-2}$s$^{-1}$. 
The geoneutrino flux is possibly of the same order of that of $^8$B solar neutrinos. 
By detecting antineutrinos from the Earth's interior, we can measure the U, Th, and K 
abundances in the Earth and their radiogenic
contribution to the heat flux. Fig. \ref{Figure:AntiNu} shows
the antineutrino spectra from the U, Th, their daughters, and K. 
Only antineutrinos in the U and Th chains have energies above 1.8 MeV, therefore being detectable
by inverse beta decay on protons. 

The KamLAND collaboration has recently presented first evidence
of geoneutrino observation~\cite{KamLAND_Geo}. The two main background sources for such 
measurement were antineutrinos from reactors and coincidence events from $(\alpha,n)$ reactions
on $^{13}$C originating from $^{210}$Po contamination in the LS (see Section 4). 

The potential of Borexino for geoneutrino detection was estimated
using the CTF3 data. 
CTF itself is too small to search for geoneutrinos. 
The analysis presented above gives nevertheless
useful information on the sensitivity potential for Borexino. 
As stated above, the low $^{210}$Po contamination makes the 
$^{13}$C-induced background in CTF negligible 
while the expected background due to $\bar{\nu}_{e}$'s
from nuclear reactors is $\sim$ 0.01 events/(ton$\times$yr) in the 1.8-3.3 MeV $\bar{\nu}_e$ 
energy window.
CTF can then set an upper limit for the geoneutrino flux. 
No candidate event was observed in the 1.8 -3.3 MeV scintillation energy range for a
7.8 ton$\times$yr exposure. The ratio between the Th and U geoneutrino fluxes can be written: 

\begin{equation}
\frac{\Phi(Th)}{\Phi(U)}=\frac{A(Th)}{A(U)}\frac{a(Th)}{a(U)}=0.83\pm0.12\label{EqGeo1}\end{equation}
where $A(Th)$ and $A(U)$ are the U and Th $\bar{\nu}_e$ specific activities, 
and $a(Th)$ and $a(U)$ are the corresponding
concentrations. Eq. (\ref{EqGeo1}) uses the value $a(Th)/a(U)=3.8\pm0.5$
from~\cite{Fogli} and allows one to express the number of expected geoneutrino events as: 

\begin{equation}
\Phi(U)=\frac{N_{geo}}{\epsilon\times N_{p}\times t\times\langle
P_{ee}\rangle\times(1+\rho)\times\langle\sigma_{U}\rangle}
\label{EqGeo2}\end{equation}
where $\langle P_{ee}\rangle=0.592\pm0.005$,
$\langle\sigma_{U}\rangle=4.24\times10^{-45}$ cm$^{2}$, 
$\rho=\frac{\Phi(Th)\langle\sigma_{Th}\rangle}{\Phi(U)\langle\sigma_{U}\rangle}=0.27\pm0.04$ 
(with $\langle\sigma_{Th}\rangle=1.30\times10^{-45}$ cm$^{2}$) and 
$\varepsilon$, $N_{p}$, $t$ are the detection efficiency, the number of target protons, and the exposure
time, respectively.

\begin{table}
\caption{Estimated background and systematics used
in geoneutrino analysis.}
\label{Table:Backgrounds}
\begin{tabular}{ll}
\hline
\bf{Backgrounds} & \bf{Expected events} \\
\hline
accidental coincidences & 0.01 \\
reactor antineutrinos & 0.11 \\
fast n,p scattering&$0.7\pm0.3$\\
\hline
\bf{Systematic uncertainties }& \% \\
\hline
efficiency, $\epsilon$ & 2 \\
number of protons, $N_{p}$ & 3.4\\
$\langle P_{ee} \rangle$ & 0.8 \\
$\rho$ & 16 \\
Energy threshold&$<2$\\
Livetime & 2 \\
\hline
\end{tabular}\end{table}

\begin{figure}
\resizebox{0.5\textwidth}{!}{%
  \includegraphics{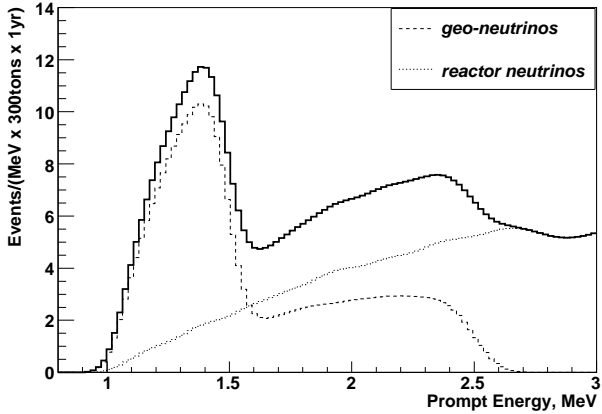}
}
\caption{The expected signal due to geoneutrinos and reactor
antineutrinos in Borexino. The simulated spectrum is normalized to the expected event rate.}
\label{Figure:Geo2}       
\end{figure}

Using the uncertainties reported in Tab.~\ref{Table:Backgrounds},
 5.2 coincidences for zero candidate events have been found for 99\% C.L.

This corresponds to an upper bound on the Uranium antineutrino flux, $\Phi(U)$, of $1.8\times10^{8}$ cm$^{-2}$s$^{-1}$.
KamLAND's current upper limit is 1$\times$10$^{7}$ (99\% C.L.) for $\Phi(U)$,
only 20 times better 
with an exposure approximately 150 times greater. The CTF result shows how important 
the purity of the LS and a low reactor background are for the detection of geoneutrinos. 
Borexino is expected to have even lower radioactive contamination than CTF
and the same specific background from reactors. Fig. \ref{Figure:Geo2}
shows the expected signal from geoneutrinos and reactor antineutrinos
in Borexino for a LS target mass of 300 tons in one year of data
taking (80\% detection efficiency). In the 1.0-2.6 MeV energy range 
5.7 events from reactor antineutrinos and 6.3 from geoneutrinos are expected,
assuming reference fluxes from \cite{Fiorentini} with corrections
for the Gran Sasso laboratory geographic position \cite{MantovaniPrivate}: 
$\Phi(U)$=4.31$\times$10$^{6}$~cm$^{-2}$s$^{-1}$
and $\Phi(Th)$=3.81$\times$10$^{6}$~cm$^{-2}$s$^{-1}$. If $S$
and $B$ are the signal and the background rates in
units of 1/(yr$\times$300 tons), $T$ is the data taking time in unit
of years, and $r=B/S$ is the background-to-signal ratio, the relative statistical 
error on the signal is: 

\begin{equation}
\delta S\equiv\frac{\Delta S}{S}=
\sqrt{\frac{1+2r}{ST}}\label{EqGeo3}\end{equation}

From Eq. (\ref{EqGeo3}) we determine that
$\delta S \sim 0.24$ in five years. The target mass limits the detection
sensitivity if the analysis is based only on rates. In a real experiment
one can perform a maximum likelihood analysis of the $\bar{\nu}_e$
energy spectrum. As shown in Fig.~\ref{Figure:Geo2}, the
geoneutrino spectrum is clearly visible over the reactor background thanks
to its distinguishing features. Both U and Th contribute to the first peak around 1.4 MeV
in Fig.~\ref{Figure:Geo2}, while only $\bar{\nu}_e$ from the U chain contribute
to the shoulder at 2.2 MeV. The U and Th contributions can thus be identified 
and better sensitivities reached.

\section{Conclusions}

The sensitivity of a high-purity liquid scintillator detector located
at the Gran Sasso Underground Laboratory to electron antineutrinos
has been investigated. The Borexino
prototype detector (CTF) was able to reach a good sensitivity
in spite of its small size compared to other liquid scintillator or
Cherenkov detectors, and set a limit for the ratio of antineutrino to neutrino fluxes from the Sun 
of $\phi_{\bar{\nu}_{e}}/\phi_{^{8}B} < 1.9\times10^{-2}$
(90\% C.L.) for $E_{\bar{\nu}} > 1.8$ MeV. The sensitivity of CTF is the result of its
very high purity from radioactive contamination combined with low reactor antineutrino background at the experimental site. The CTF data also show that Borexino
can search for electron antineutrinos
from the Sun and from the Earth interior with very competitive sensitivity. 
In particular, the expected
sensitivity to a possible solar $\bar{\nu}_e$ flux is at the level of $\phi_{\bar{\nu}_{e}}/\phi_{^{8}B} \sim 10^{-5}$, which is of interest for both looking for a neutrino magnetic moment and for studying the magnetic field inside the Sun.

\begin{acknowledgement}
We would like to thank F.Mantovani and A.Palazzo for useful discussions
during the preparation of the paper.
\end{acknowledgement}


\end{document}